\documentclass[amsmath,pra,twocolumn,showpacs]{revtex4}
\usepackage{graphics}

\begin{document}     

\title{Atom trapping and guiding with a subwavelength-diameter optical fiber}
\date{\today}

\author{V. I. Balykin,$^{1,2,4}$ K. Hakuta,$^{1,3}$ Fam Le Kien,$^{1,3,*}$ J. Q. Liang,$^{1,3}$ and M. Morinaga$^{1,2}$} 
\affiliation{$^1$Course of Coherent Optical Science, 
University of Electro-Communications, Chofu, Tokyo 182-8585, Japan\\
$^2$Institute for Laser Science, University of Electro-Communications, Chofu, Tokyo 182-8585, Japan\\
$^3$Department of Applied Physics and Chemistry, 
University of Electro-Communications, Chofu, Tokyo 182-8585, Japan\\
$^4$Institute of Spectroscopy, Troitsk, Moscow Region, 142092, Russia}

\date{\today}

\begin{abstract}

We suggest using an evanescent wave around a thin fiber to trap atoms. 
We show that  the gradient force
of a red-detuned evanescent-wave field in the fundamental mode of a silica fiber can balance the centrifugal force
when the fiber diameter is about two times smaller than the  wavelength of the light 
and the component of the angular momentum of the atoms  along the fiber axis 
is in an appropriate range.  As an example, the system should be  realizable for Cesium atoms at a temperature of
less than  0.29 mK using
a silica fiber with a radius of 0.2 $\mu$m and a  1.3-$\mu$m-wavelength  light
with a power of about 27 mW.

\end{abstract}

\maketitle

\begin{figure}
\begin{center}
  \includegraphics{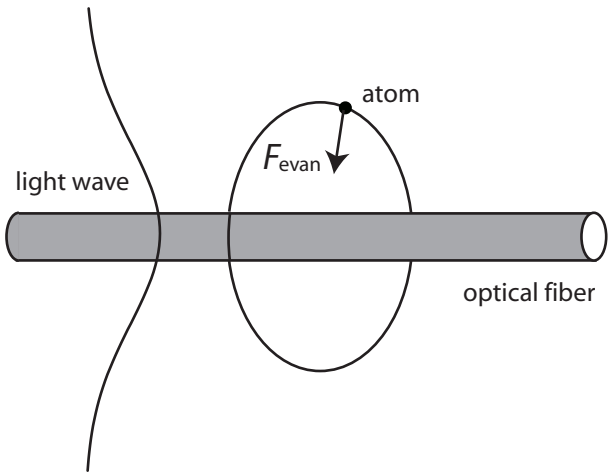}
 \end{center}
\caption{Schematic of atom trapping and guiding around an optical fiber.}
\label{fig1}
\end{figure}

A current-carrying metallic wire and a photon-carrying dielectric wire are the simplest configurations for trapping and guiding  cold particles. Blumel and Dietrich \cite{b1} considered the possibility of binding very cold neutrons through a magnetic trapping potential created by a thin wire with a current. Similar waveguides have been proposed and lately realized  for binding cold neutral atoms \cite{b2,b3}. The optical force from a dielectric wire carrying photons can also be used to trap and guide cold atoms \cite{b4}. Such an atom guide is made from a \textit{hollow} optical fiber with light propagating in the glass and tuned far to blue of atomic resonance \cite{b5,b6}.   Inside  the fiber, the evanescent wave decays exponentially away from the wall, producing a repulsive potential which guides atoms along the center axis. Alternatively, a red-detuned light in the hollow center of the fiber can also be used to guide atoms \cite{b3}. 
In several experiments  \cite{b7}, cold atoms have been trapped and guided \textit{inside} a hollow fiber where  the optical dipole force confined them on axis.

In this Rapid Communication, we present a  method for trapping and guiding  neutral atoms 
around a thin optical fiber. Our scheme is based on the use of a \textit{subwavelength-diameter} silica fiber with a red-detuned light launched into it. The light wave decays away from the fiber wall and produces an attractive potential for  neutral atoms. The atom trapping and guiding occur in the \textit{outside} of  the fiber. To sustain a stable trapping and guiding, the atoms have to be kept away from the fiber wall. This can be achieved by the centrifugal potential barrier.  The centrifugal potential barrier can compensate all the potentials which diverge less rapidly than $r^{-2}$ as $r \to 0$. We show that this can be achieved only when the fiber diameter  is  subwavelength.  (Nowadays thin fibers can be produced with diameters down to 50 nm \cite{b8}.)

Note that one can manipulate a very thin silica fiber using taper fiber technology. 
The essence of the technology is to heat and 
pull a single-mode optical fiber to a very thin thickness 
maintaining the taper condition to keep adiabatically 
the single-mode condition \cite{taper}. The thin 
fiber has an almost vanishing core.  
Hence, the refractive
indices that determine the fiber modes are the cladding refractive index $n_f$ and the refractive index $n_0$
of the medium surrounding the fiber. The cladding index $n_f$ will be referred as the fiber refractive index
henceforth.

Consider an atom  moving around an optical fiber, see Fig. \ref{fig1}. 
We assume that the potential $U$ of the atom is cylindrically symmetric, that is, $U$ depends on  
the radial distance $r$ from the atom to the fiber axis $z$, but not on two other cylindrical coordinates $\varphi$ and $z$. Due to this symmetry, the component $L_z$ of the angular momentum of the atom is conserved. 
In the eigenstate problem, we have $L_z=\hbar m$, where $m$ is an integer, called the rotational quantum number.
The centrifugal potential of the atom is repulsive and  is given by $U_{\mathrm{cf}}={\hbar^2(m^2-1/4)}/{2M r^2}$. The  radial motion of the atom can be treated as the one-dimensional motion of a particle in the effective potential $U_{\mathrm{eff}}=U_{\mathrm{cf}}+ U$.

There exist stable bound states for the atom if
$U_{\mathrm{eff}}$ has a local minimum at a  distance $r=r_m$ outside the fiber. This may happen only if
the potential $U$ is attractive, opposite to the centrifugal potential $U_{\mathrm{cf}}$. 
To produce such a potential $U$, we send an optical field through the fiber. This field generates an evanescent wave around the fiber, whose
steep variation in the transverse plane leads to a gradient force on the atom. 
We assume that the atom is initially in the ground state and
the  detuning $\Delta$ of the field from the dominant atomic line
is large compared to the Rabi frequency $\Omega$ and the linewidth $\gamma$. 
Then, the optical potential of the gradient force is  given by \cite{Kazantsev,Balykin}
$U=\hbar\Omega^2/\Delta$.
We choose a red detuning ($\Delta<0$) for the field to make $U$ an attractive potential.

We assume that the fiber is sufficiently thin that it has a vanishing core and it can support only a single, fundamental mode HE$_{11}$ \cite{fiber books}. 
The fiber mode characteristics are then determined by the fiber radius $a$, the light wavelength $\lambda$, the fiber refractive index $n_f$, and the refractive index $n_0$
of the surrounding environment. 
In the linear-polarization approximation,  
the spatial dependence of the amplitude of the field outside the fiber is described by  
the modified Bessel function $K_0(qr)$.
Here the parameter $q=1/\Lambda$ is the inverse of 
the characteristic decay length $\Lambda$ of the evanescent-wave field 
and is determined by the fiber eigenvalue equation \cite{fiber books}. 
Then,  the optical potential outside the fiber can be written as  
$U=-GK_0^2(qr)$, where $G=-\hbar\Omega_a^2/\Delta K_0^2(qa)$ is the coupling constant for the interaction between the evanescent wave and the atom. Here $\Omega_a$ is the Rabi frequency of the field at the fiber surface.

It should be noted here that the field distribution $\mathbf{E}(\mathbf{r})$ corresponding to the fundamental mode HE$_{11}$ of the fiber has
three nonzero components $E_r$, $E_{\varphi}$, and $E_z$, which have azimuthal variation 
and therefore are not cylindrically symmetric. 
The cylindrical symmetry of the field in the fundamental mode appears only in the framework
of the linear-polarization approximation for weakly guided modes. Although this approximation is  good
for conventional fibers,  it may be questionable for thin fibers since $n_f$ and $n_0$ are very different from each other.
A simple general way to produce a cylindrically symmetric optical potential is to use  a circularly polarized light.
The time average of the potential of such a  field is cylindrically 
symmetric on the slow time scale of the atomic center-of-mass motion.

The effective potential for the radial motion of the atom in the optical potential $U$ 
can be written in the form
\begin{equation}
U_{\mathrm{eff}}(r)=\theta_{\mathrm{rec}}\left[\frac{m^2-1/4}{ k^2r^2}-g K_0^2(q r)\right],
\label{12d}
\end{equation}
where $\theta_{\mathrm{rec}}=(\hbar k)^2/2M$ is the recoil energy and $g=G/\theta_{\mathrm{rec}}$ is the normalized coupling parameter. 
Here $M$ is the mass of the atom and $k$ is the wave number of the field.
At the local minimum point $r_m$, the derivative of $U_{\mathrm{eff}}(r)$ is zero. 
Hence,  we find $r_m=x_m/q$, where $x_m$ is a solution of the equation 
\begin{equation}
f(x)={\cal M}, 
\label{7}
\end{equation}
with
$f(x)=x^3 K_0(x)K_1(x)$ and ${\cal M}=(m^2-1/4){q^2}/{gk^2}$.
The function $f(x)$ achieves its peak value $f_c\cong0.2545$ at $x_c\cong0.9331$. 
The condition for the existence of  $r_m$ is ${\cal M}< f_c$.
When this condition  is satisfied, Eq. (\ref{7}) has two solutions. 
The smaller one  corresponds to the local minimum point $r_m$ of $U_{\mathrm{eff}}(r)$
and the larger one corresponds to a local maximum point.
It follows from the relation $x_m<x_c$ that
$r_m<x_c\Lambda\cong0.9331\, \Lambda$.
Thus the distance from the local minimum point $r_m$ of the effective optical  potential 
to the fiber axis is always shorter than the decay length $\Lambda$ of the evanescent wave.
The comparison of the requirement $r_m>a$ with the inequality $r_m<x_c\Lambda$ yields
\begin{equation}
q a<x_c\cong0.9331
\label{12}
\end{equation}
or, equivalently, $a<x_c\Lambda\cong0.9331\,\Lambda$. As shown below,
the  condition (\ref{12}) can be satisfied only when the size parameter $ka$ of the fiber is small enough.
On the other hand, the requirement $r_m>a$ can be satisfied only when
$f_a<{\cal M}$, where $f_a=f(x_a)$ with $x_a=qa$.  
Combining the conditions $f_a<{\cal M}$ and ${\cal M}< f_c$ and using the explicit expression for ${\cal M}$, we obtain 
\begin{equation}
f_a<(m^2-1/4)\frac{q^2}{gk^2}<f_c.
\label{12e}
\end{equation}
Thus   stable bound states of the atom may exist only if 
$m_{\mathrm{min}}\leq m\leq m_{\mathrm{max}}$,
where $m_{\mathrm{min}}$ and $m_{\mathrm{max}}$ are the integer
numbers closest to $\sqrt{f_agk^2/q^2+1/4}$ and $\sqrt{f_cgk^2/q^2+1/4}$, respectively.
Note that an increase in $m$ leads to an increase in the position $r_m$ as well as a decrease in the depth $-U_{\mathrm{eff}}(r_m)$ of the local minimum of the effective optical  potential.
In addition, the depth $-U_{\mathrm{eff}}(r_m)$  increases with  increasing  the coupling parameter $g$.

\begin{figure}
\begin{center}
  \includegraphics{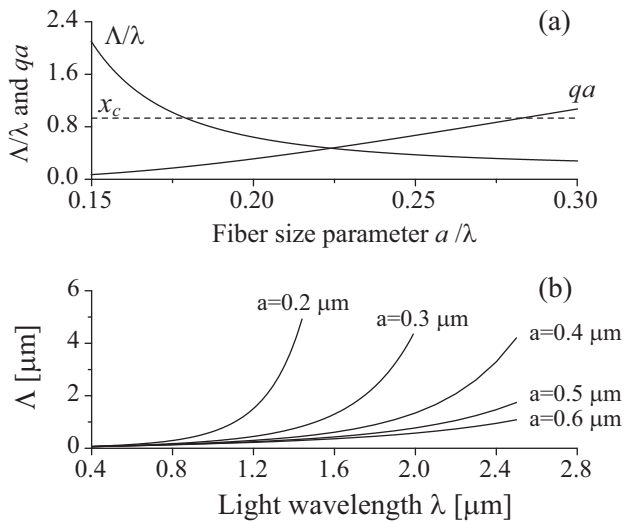}
 \end{center}
\caption{(a) Normalized penetration depth $\Lambda/\lambda$ and normalized decay rate $q a$ of the evanescent wave  as functions of the fiber size parameter $a/\lambda$.
The parameters used: $n_f=1.45$ and $n_0=1$.
(b) Penetration length $\Lambda$ of the evanescent wave against the wavelength $\lambda$ of the light field
for various values of the fiber radius $a$.
The dispersion of the silica-glass fiber  index $n_f$ is taken into account, and the environment refractive index
is $n_0=1$.
}
\label{fig2}
\end{figure}

We calculate the  decay parameter $q$ for the evanescent wave of the fundamental mode HE$_{11}$ by
solving the exact eigenvalue equation for the fiber  \cite{fiber books}.
We show in Fig.~\ref{fig2}(a) the characteristic dimensionless parameters $\Lambda/\lambda$ and  $q a$ 
as  functions of the fiber size parameter $a/\lambda$. 
We find from the figure that the condition  (\ref{12}) is satisfied
when 
\begin{equation}
a/\lambda < 0.283, 
\label{c1}
\end{equation}
that is, when the fiber is thin compared to the wavelength of the trapping light.
The penetration length $\Lambda$ is plotted in Fig. \ref{fig2}(b)
as a function of the light wavelength $\lambda$ for various values of the fiber radius $a$.
In these calculations, we took into account the dispersion of the silica-glass fiber  index $n_f$. 
The figure shows that the condition (\ref{12}) is satisfied
when $\lambda> 0.72$, 1.06, 1.41, 1.75, and 2.09 $\mu$m for $a=0.2$, 0.3, 0.4, 0.5, and 0.6 $\mu$m,
respectively. 

\begin{figure}
\begin{center}
  \includegraphics{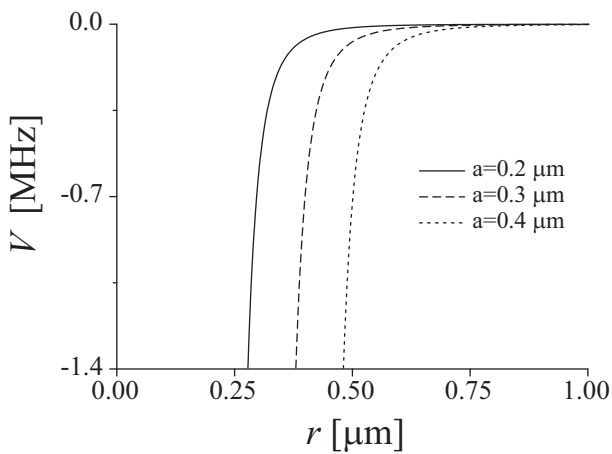}
 \end{center}
\caption{van der Waals potential $V$ of a ground-state cesium atom in the outside of
a thin cylindrical silica fiber. 
}
\label{fig3}
\end{figure}

When the atom is near to the fiber surface, we must take into account the van der Waals force.
The van der Waals potential of an atom near the surface of a cylindrical dielectric rod 
has been calculated by Boustimi \textit{et al.} \cite{Baudon}.
We use their general formula to calculate the van der Waals potential of a ground-state cesium atom near a cylindrical silica fiber. We plot the results in Fig. \ref{fig3}.
As seen, when $r$ is fixed, a larger fiber radius $a$ leads to a deeper and steeper van der Waals potential. 
When the van der Waals force is substantial, the effect of the centrifugal potential may be reduced and, consequently,  
the local minimum point $r_m$ of the effective optical potential $U_{\mathrm{eff}}$ may be washed out
in the total effective potential $U_{\mathrm{tot}}=U_{\mathrm{eff}}+V$. To reduce the effect of the van der Waals force, we need a thin fiber.
Around such a fiber, 
the total effective potential $U_{\mathrm{tot}}$ may have a local minimum point near to $r_m$.
On the other hand, we can also reduce the effect of the van der Waals force by increasing $r_m$. 
For a fixed value of the coupling parameter $g$, we can increase
$r_m$ by considering orbits with  larger values of the rotational quantum number $m$.

We now perform numerical calculations for the total effective potential $U_{\mathrm{tot}}$ 
of a ground-state cesium atom in the outside of a thin cylindrical silica fiber.
To trap atoms, we use light with the wavelength $\lambda=1.3$ $\mu$m.
The fiber radius is chosen to be $a=0.2$ $\mu$m $=0.154\,\lambda$, which is  small enough 
to satisfy the condition (\ref{c1}) and to reduce the effect of the van der Waals force. 
The penetration length of the trapping light from the fiber is found to be $\Lambda\cong 2.42$ $\mu$m $=12.1\,a$.
The coupling parameter $g$ should be large to produce a strong dipole force 
leading to an effective optical
potential with a deep local minimum point.
For the calculations, we choose $g=5330$.
Then, the condition (\ref{12e}) yields  the lower and upper boundary
values $m_{\mathrm{min}}=114$ and $m_{\mathrm{max}}=430$ for the range of $m$
where the effective optical $U_{\mathrm{eff}}$ has a local minimum point $r_m$ outside the fiber.
For the calculations, we choose $m=220$, 230, and 240. These values are  not only in the interval 
$(m_{\mathrm{min}},m_{\mathrm{max}})$ 
but  also are sufficiently large that $r_m$ is far away from the range of substantial action of the van der Waals force. 
We then expect that the total effective potential $U_{\mathrm{tot}}$
has a local minimum point near to the point  $r_m$.

\begin{figure}
\begin{center}
  \includegraphics{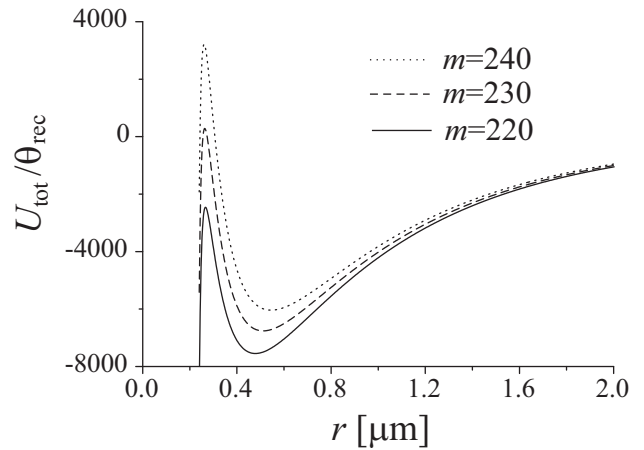}
 \end{center}
\caption{
Total effective potential $U_{\mathrm{tot}}$ 
 of a ground-state cesium atom in the outside of
a silica fiber with the radius of 0.2 $\mu$m. The wavelength and the coupling parameter of the trapping light
are $\lambda=1.3$ $\mu$m and  $g=5330$, respectively. 
The recoil energy is $\theta_{\mathrm{rec}}=888$ Hz $=42.6$ nK. 
} 
\label{fig4}
\end{figure}

We plot in Fig. \ref{fig4} the results of our calculations for  $U_{\mathrm{tot}}$.
As seen, the total effective potential has a deep minimum point, which is located more than 0.4 $\mu$m far away from the fiber axis,
not only well  outside the fiber
but also outside the range of substantial action of the van der Waals force. 
We observe that an increase in the rotational quantum number $m$ 
leads to an increase in the position $r'_m$ of the local minimum point 
and a decrease in its depth $-U_{\mathrm{tot}}(r'_m)$.
We note that, in the region of  $r\simeq a$, the shape of $U_{\mathrm{tot}}$ is similar to that of the attractive van der Waals potential $V$. 
However, in the region of  $r\gg a$, where $V$ is weak,
$U_{\mathrm{tot}}$ practically coincides with $U_{\mathrm{eff}}$.

\begin{figure}
\begin{center}
  \includegraphics{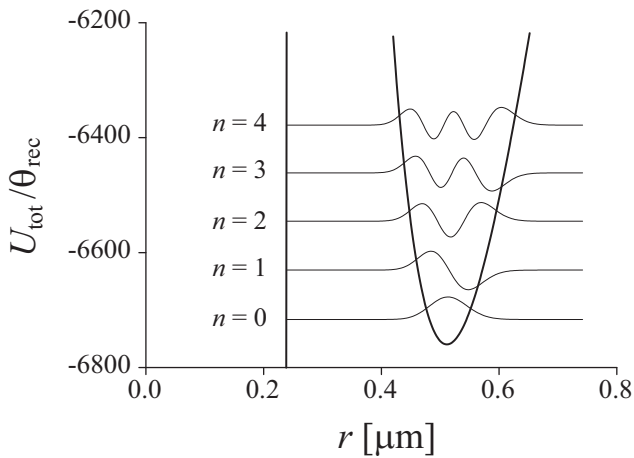}
 \end{center}
\caption{Bound states $u_n$ for the first five levels ($n=0$, 1, 2, 3, and 4) of the radial motion a cesium atom in the total effective potential $U_{\mathrm{tot}}$.
The rotational quantum number is $m=230$.
All the other parameters are the same as for Fig. \ref{fig4}.
} 
\label{fig5}
\end{figure}

The existence of a deep local minimum of the  potential $U_{\mathrm{tot}}$ leads
to the existence of bound states of the atom.
In Fig. \ref{fig5}, we plot the wave functions $u_n$ for the first five energy levels 
of the radial motion of  a cesium atom in the  potential $U_{\mathrm{tot}}$. Since the potential is deep, the lower levels practically do not depend 
on the van der Waals potential. They are mainly determined by the effective optical potential
$U_{\mathrm{eff}}$. However, the upper levels, which are not shown in the figure, are sensitive
to the van der Waals potential and also to the boundary condition at the fiber surface.
Tunneling from such highly excited levels into the narrow potential well between the repulsive hard-core  potential and the attractive van der Waals potential may occur.

The experimental realization of bound states requires the  binding energy of atoms to be larger
than their typical thermal kinetic energy. In the case of Fig. \ref{fig5}, the energy of the  ground state
is $E_0\cong-6716\,\theta_{\mathrm{rec}}$. Since $U_{\mathrm{tot}}(r=\infty)=0$,
the binding energy of an atom in the ground state is $E_B=U_{\mathrm{tot}}(r=\infty)-E_0 
\cong6716\,\theta_{\mathrm{rec}}\cong0.29$ mK. 
Thus we can trap cesium atoms around the fiber at a temperature of less than 0.29 mK.

The tangential component of transverse velocity of an atom spinning around the fiber at the distance $r_m$ is given by $v_T=L_z/Mr_m=\hbar m/Mr_m$.
For a cesium atom spinning in a bound state with $m=230$ and $r_m=0.5$ $\mu$m, we find $v_T\cong 0.22$ m/s.
The  kinetic energy of this rotational motion is $Mv_T^2/2\cong 1$ mK. This energy is larger than the binding energy and consequently than the allowed value of the average thermal kinetic energy  estimated above.

We note that the coupling parameter $g=5330$ 
corresponds to  the field intensity  of 1 MW/cm$^2$  at the fiber surface with the radius
of 0.2 $\mu$m.
Such an intensity is large, but it can be easily generated since the fiber is thin.
In fact, the power of light corresponding to  such an intensity is estimated to be about 27 mW,
a very ordinary value for the laser light with the wavelength of 1.3 $\mu$m.

In conclusions, we have suggested and analyzed a method for trapping atoms based on the use of an evanescent wave around a thin silica fiber. 
We have shown that  the gradient force
of a red-detuned evanescent-wave field in the fundamental mode of a silica fiber can balance the centrifugal force
when the fiber diameter is about two times smaller than the  light wavelength and the component of the angular momentum of the atoms  along the fiber axis is in an appropriate range.
Since optical fibers offer the possibility of engineering evanescent fields, the system can be used
to store,  move, and manipulate cold atoms in a controlled manner.
The system may prove useful as a waveguide for matter waves.


\begin{thebibliography}{99}


\bibitem[$*$]{a}  Also at Institute of Physics, National Center for 
Natural Sciences and Technology, Hanoi, Vietnam. 


\bibitem{b1} R. Blumel and K. Dietrich, Phys. Rev. A \textbf{43}, 22 (1991).

\bibitem{b2} L.V. Hau, J.A. Golovchenko, and M.M. Burns, Phys. Rev. Lett. \textbf{74}, 3138 (1995).

\bibitem{b3}  J. Schmiedmayer,  Phys. Rev. A \textbf{52}, R13 (1995); Appl. Phys. \textbf{B 60}, 169 (1995).


\bibitem{b4} M.A. Ol'Shanii, Yu.B. Ovchinnikov, and V.S. Letokhov, Opt. Commun. \textbf{98}, 77 (1993). 

\bibitem{b5} M.J. Renn, D. Montgomery, O. Vdovin, D.Z. Anderson, C.E. Wieman, and  E.A. Cornell, 
Phys. Rev. Lett. \textbf{75}, 3253 (1995).

\bibitem{b6} H. Ito, T. Nakata, K. Sakaki, M. Ohtsu, K. I. Lee,  and W. Jhe,  Phys.  Rev. Lett. \textbf{76}, 4500 (1996). 

\bibitem{b7} For review see, for example, J.P. Dowling and J. Gea-Banacloche, 
Adv. At. Mol. Opt. Phys. \textbf{37}, 1 (1996); V. I. Balykin, Adv. At. Mol. Opt. Phys. \textbf{41}, 181 (1999).

\bibitem{b8} L. Tong, R.R. Gattass, J.B. Ashcom, S. He, J. Lou, M. Shen, I. Maxwell, and E. Mazur, Nature \textbf{426}, 816 (2003).


\bibitem{taper} J.C. Knight, G. Cheung, F. Jacques, and T.A. Birks, Opt. Lett. \textbf{22}, 1129 (1997);
T.A. Birks, W.J. Wadsworth, and P. St. J. Russell,  Opt. Lett. \textbf{25}, 1415 (2000); 
M. Cai and K. Vahala, Opt. Lett. \textbf{26}, 884 (2001).

\bibitem{Kazantsev} A.P. Kazantsev, G.J. Surdutovich, and V.P. Yakovlev, \textit{Mechanical Action of Light on Atoms} (World Scientific, Singapore, 1990).

\bibitem{Balykin} V.I. Balykin, V.G. Minogin, and V.S. Letokhov,
Rep. Prog. Phys. \textbf{63}, 1429 (2000). 

\bibitem{fiber books} See, for example, A. Yariv, \textit{Optical Electronics} (CBS College Publishing, New York, 1985); 
D. Marcuse, \textit{Light Transmission Optics} 
(R. E. Krieger Publishing Co., Malabar, Florida,  1989).

\bibitem{Baudon} M. Boustimi, J. Baudon, P. Candori, and J. Robert,
Phys. Rev. B \textbf{65}, 155402 (2002); M. Boustimi, J. Baudon, and J. Robert, Phys. Rev. B \textbf{67}, 045407 (2003).  

\end{thebibliography}
\end{document}